\documentclass{article}
\usepackage{spconf,amsmath,graphicx}

\usepackage{amsmath,amssymb,amsfonts}
\usepackage{multirow}
\usepackage{color,soul}
\usepackage{algorithm,algorithmic}
\usepackage{multirow}
\DeclareMathOperator{\softmax}{softmax}
\usepackage{enumitem}
\setlist{nosep, leftmargin=14pt}

\usepackage{mwe} 

\title{KOCOBrain: Kuramoto-Guided Graph Network for Uncovering Structure-Function Coupling in Adolescent Prenatal Drug Exposure}
\name{Badhan Mazumder$^{\star \dagger}$ \quad Lei Wu$^{\dagger}$ \quad Sir-Lord Wiafe$^{\star \dagger}$ \quad Vince D. Calhoun$^{\star \dagger}$ \quad Dong Hye Ye$^{\star \dagger}$}
  
\address{$^{\star}$ Department of Computer Science, Georgia State University, Atlanta, GA, USA \\
      $^{\dagger}$ Tri-Institutional Center for Translational Research in Neuroimaging and Data Science (TReNDS),\\ Georgia State University, Georgia Institute of Technology, and Emory University, Atlanta, GA, USA}
\makeatletter
\let\oldthebibliography\thebibliography
\def\thebibliography#1{%
  \oldthebibliography{#1}%
  \setlength{\itemsep}{0pt}%
  \setlength{\parskip}{0pt}%
  \setlength{\parsep}{0pt}%
}
\makeatother
\begin{document}
%
\maketitle
\begin{abstract}
Exposure to psychoactive substances during pregnancy, such as cannabis, can disrupt neurodevelopment and alter large-scale brain networks, yet identifying their neural signatures remains challenging. We introduced KOCOBrain: KuramotO COupled Brain Graph Network; a unified graph neural network framework that integrates structural and functional connectomes via Kuramoto-based phase dynamics and cognition-aware attention. The Kuramoto layer models neural synchronization over anatomical connections, generating phase-informed embeddings that capture structure–function coupling, while cognitive scores modulate information routing in a subject-specific manner followed by a joint objective enhancing robustness under class imbalance scenario. Applied to the ABCD cohort, KOCOBrain improved prenatal drug exposure prediction over relevant baselines and revealed interpretable structure-function patterns that reflect disrupted brain network coordination associated with early exposure.

\end{abstract}
\vspace{-12pt}
\section{Introduction}
\vspace{-10pt}
Maternal use of psychoactive substances such as cannabis during pregnancy has been increasingly recognized as a risk factor for long-term neurodevelopmental alterations \cite{I_2,My_1}. Adolescence, a period marked by rapid remodeling of large-scale brain networks that support cognition, emotion regulation, and executive control \cite{I_0} often reveals the latent impact of these early exposures. Recent neuroimaging studies suggest that prenatal drug exposure (PDE) can perturb both the integrity of white-matter tracts and the coordination of functional networks \cite{My_1,Qual_3}, yet the precise neural mechanisms underlying these disruptions remain unclear. Reliable identification of such neurological signatures is critical for advancing early risk detection and understanding how early perturbations shape the developing connectome. 

Despite rapid progress in computational neuroimaging research domain , current modeling approaches face three persistent challenges. First, many studies \cite{base3} rely on either structural or functional connectivity alone, overlooking their mutual dependence. Structural connectivity (SC) from diffusion tensor imaging (DTI) provides the physical wiring for neural communication, while functional network connectivity (FNC) from resting-state functional MRI (rs-fMRI) captures dynamic interactions that emerge from it; ignoring their coupling can obscure biologically meaningful information. 
Second, most existing graph neural network (GNN) based methods \cite{base5,base6} represent these connectomes as static entities, lacking mechanisms to capture the oscillatory synchronization that underlies brain coordination. Such simplifications limit their ability to model the nonlinear dynamics linking anatomy and function.  Third, individual cognitive variability, which is considered an important factor \cite{My_1} influencing how network interactions manifest, is often omitted or only weakly incorporated \cite{My_2,My_3}. This neglect can mask subject-specific adaptations, especially in heterogeneous developmental populations such as PDE cohorts.

To address these limitations, we proposed KOCOBrain: KuramotO COupled Brain Graph Network that unifies structural and functional connectomes within a biologically informed, cognition-aware GNN framework. KOCOBrain employs a differentiable Kuramoto phase module constrained by each subject’s SC, simulating neural synchronization to capture structure-function coupling. These phase representations, combined with local structural topology, form graph node tokens processed through cognition-gated attention layers according to individual cognitive profiles. Anatomical masking ensures biologically plausible communication pathways, while a joint learning objective enhances stability under class imbalance scenario in an end-to-end manner. 

\begin{figure*}[htb]
  \centering
  \centerline{\includegraphics[width=\linewidth]{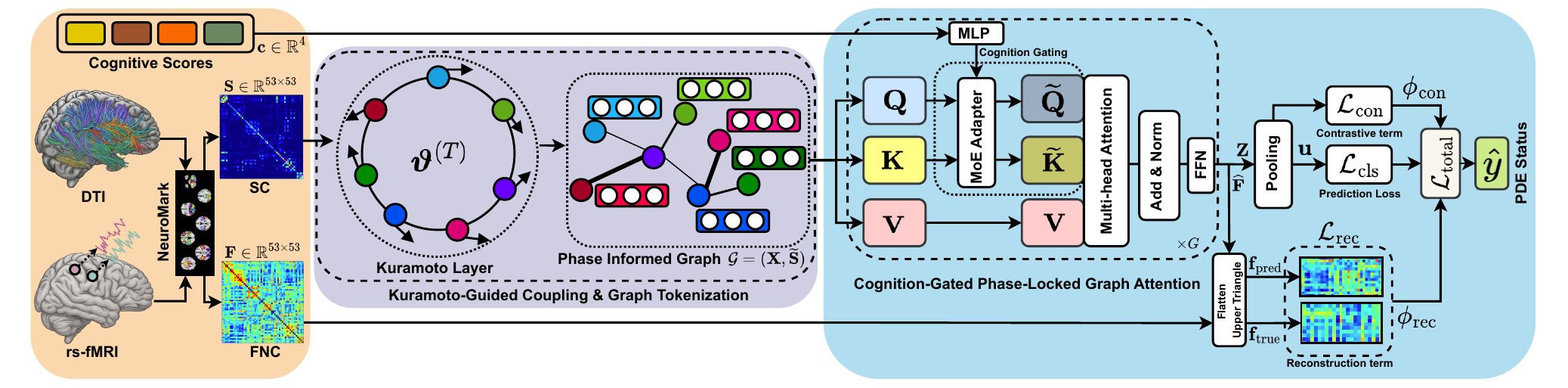}}
  \vspace{-10pt}
\caption{\textbf{KOCOBrain overview:} SC from DTI and FNC from rs-fMRI were jointly modeled via a Kuramoto-guided coupling layer that simulated oscillatory phase synchronization over the anatomical wiring. The resulting phase dynamics yielded structure-informed functional representations, which combined with local structural features formed phase-aware node tokens. These tokens passed through cognition-gated, phase-locked graph-transformer layers where cognitive scores modulated query-key interactions via low-rank MoE adapters, and attention weights were biased by phase synchrony and SC strength. Global pooling of the refined embeddings produced a subject-level representation optimized end-to-end for PDE status prediction.}
\vspace{-13pt}
\label{fig:fig1}
\end{figure*}
The main contributions of our work are threefold: 

\begin{enumerate}
    \item We introduced a differentiable Kuramoto-guided coupling module that modeled phase synchronization directly over each subject’s SC, providing a biologically grounded mechanism for linking structure and function. 
    \item We developed a cognition-gated graph attention mechanism that integrated subject-level cognitive measures to modulate information routing across the network. 
    \item We designed a joint optimization strategy that enhanced stability, interpretability, and robustness in predicting PDE status. 

\end{enumerate}

\section{Methodology}
\vspace{-8pt}
\noindent
Let \(\mathcal{D} = \{(\mathbf{S}_m, \mathbf{F}_m, \mathbf{c}_m, y_m)\}_{m=1}^{M}\) denote a dataset of \(M\) subjects. For each subject \(m\), the connectome is defined over \(N = 53\) brain networks, with SC matrix \(\mathbf{S} \in \mathbb{R}^{N \times N}\) derived from DTI and FNC matrix \(\mathbf{F} \in \mathbb{R}^{N \times N}\) computed from rs-fMRI. A cognition vector \(\mathbf{c} \in \mathbb{R}^{4}\) encodes standardized scores for working memory, fluid, crystallized, and total intelligence, and \(y \in \{0,1\}\) indicates the PDE label. As shown in Fig. \ref{fig:fig1}, the objective is to learn a mapping \((\mathbf{S}, \mathbf{F}, \mathbf{c}) \rightarrow \hat{y} \in [0,1]\) that integrates multimodal connectomes and cognitive context within a unified framework for graph-level prediction.

\vspace{-10pt}
\subsection{Kuramoto-Guided Coupling \& Graph Tokenization}
\vspace{-5pt}
\subsubsection{Kuramoto Dynamics for Structure--Function Coupling}
\vspace{-5pt}
Neural communication occurs through oscillatory synchronization along anatomical wiring. To capture this coupling, we employed a differentiable Kuramoto \cite{M_0} module that simulates phase evolution over each subject’s SC, providing a dynamic, trainable link between structure and function beyond static correlations.

For a connectome of \(N\) brain networks, each node \(i\) maintains a phase \(\vartheta_i^{(t)}\) at iteration \(t\), updated as below: 
\vspace{-8pt}
\begin{equation}
\vartheta_i^{(t+1)} = \varpi\!\left(\vartheta_i^{(t)} + \Delta t \left[ \omega_i + \sum_{j=1}^{N} \widetilde{S}_{ij}\,\sin(\vartheta_j^{(t)} - \vartheta_i^{(t)}) \right] \right)
\label{eq:kuramoto_update}
\end{equation}
where \(\boldsymbol{\omega}\in\mathbb{R}^{N}\) are trainable intrinsic frequencies shared across subjects, \(\widetilde{\mathbf{S}}\) is the row-normalized SC constrained by the Top-\(k\) anatomical mask \(\mathbf{A}_{\text{SC}}\in\{0,1\}^{N\times N}\) preserving the strongest SC links per node (including self-connections), and \(\varpi(\cdot)\) wraps the phase into \((-\pi,\pi]\). The evolution begins from \(\boldsymbol{\vartheta}^{(0)}=\mathbf{0}\) and proceeds for \(T\) discrete integration steps with step size \(\Delta t\). To ensure numerical stability, the coupling term was passed through a smooth saturating nonlinearity such as \(\tanh\) before scaling. The resulting final state \(\boldsymbol{\vartheta}^{(T)}\) encoded network-level phase synchrony driven by anatomical coupling, forming a biologically interpretable bridge between structural constraints and functional coordination.
\vspace{-10pt}
\subsubsection{Phase-Informed Graph Tokenization}
\vspace{-5pt}
The final phases $\boldsymbol{\vartheta}^{(T)}$ captured network coordination and were embedded as compact node tokens combining dynamic phase and local structural features for downstream attention-based learning. Specifically, node \(i\) is represented as $\mathbf{x}_i = \big[\cos\vartheta_i^{(T)},\, \sin\vartheta_i^{(T)},\, \rho(S_{i:}),\, d_i\big]\,\mathbf{W}_x$,
where \(\rho:\mathbb{R}^{N}\!\rightarrow\!\mathbb{R}^{d_\rho}\) is an MLP summarizing the \(i\)-th structural row, \(d_i=\sum_j S_{ij}\) is the node degree, and \(\mathbf{W}_x \in \mathbb{R}^{(2 + d_\rho + 1)\times d_x}\) is a learnable projection mapping features into a \(d_x\)-dimensional token space. The trigonometric embedding of phases preserves circular geometry while stabilizing optimization, and the structural statistics provide anatomical context.  

Edges in the graph were defined by the anatomical mask \(\mathbf{A}_{\text{SC}}\), where an edge \((i,j)\) exists if \([\mathbf{A}_{\text{SC}}]_{ij}=1\). Each edge was assigned a normalized weight \(\widetilde{S}_{ij}\), ensuring that message propagation and attention were restricted to anatomically valid connections. The resulting phase-informed graph is represented as \(\mathcal{G}=(\mathbf{X},\widetilde{\mathbf{S}})\), which encodes both dynamic synchrony and structural topology.

\vspace{-10pt}
\subsection{Cognition-Gated Phase-Locked Graph Attention}
\vspace{-5pt}
Given the graph \(\mathcal{G}=(\mathbf{X},\widetilde{\mathbf{S}})\), KOCOBrain performs global self-attention under anatomical constraints, functioning as a graph transformer where attention weights are shaped by SC, phase synchrony, and cognition. The normalized structural matrix \(\widetilde{\mathbf{S}}\) provides weighted topology, while its binary support \(\mathbf{A}_{\text{SC}}\) restricts attention to anatomically relevant connections.

\textit{Base projections.} Node features \(\mathbf{X}\in\mathbb{R}^{N\times d_x}\) were projected to queries, keys, and values \cite{M_1} as
\(
\mathbf{Q}=\mathbf{X}\mathbf{W}_Q,\quad
\mathbf{K}=\mathbf{X}\mathbf{W}_K,\quad
\mathbf{V}=\mathbf{X}\mathbf{W}_V,
\)
where \(\mathbf{W}_Q,\mathbf{W}_K,\mathbf{W}_V\in\mathbb{R}^{d_x\times d_x}\) are shared across subjects. Each projection was divided into \(H\) attention heads with per-head dimension \(d_h=d_x/H\).

\textit{Cognition gating.} Cognitive variability influences how networks align and exchange information. To model this effect, queries and keys were adapted through a cognition-gated Mixture-of-Experts (MoE) adapter \cite{M_2} conditioned on the cognition vector \(\mathbf{c}\in\mathbb{R}^{4}\). A softmax gate \(\boldsymbol{\pi}(\mathbf{c})=\softmax(\mathbf{W}_\pi\psi(\mathbf{c}))\), where \(\psi(\cdot)\) is an MLP that projected \(\mathbf{c}\) into a latent space of dimension \(d_g\), and \(\mathbf{W}_\pi\in\mathbb{R}^{E\times d_g}\), mixed \(E\) low-rank experts. Each expert \(e\) was parameterized by rank-\(r_e\) factors \(\mathbf{U}^{(e)}_{Q},\mathbf{U}^{(e)}_{K}\in\mathbb{R}^{d_x\times r_e}\) and \(\mathbf{P}^{(e)}_{Q},\mathbf{P}^{(e)}_{K}\in\mathbb{R}^{r_e\times d_x}\), yielding the adapted projections:
\vspace{-5pt}
\begin{equation}
\widetilde{\mathbf{Q}}=\mathbf{Q}+\sum_{e=1}^{E}\pi_e(\mathbf{c})(\mathbf{X}\mathbf{U}^{(e)}_{Q})\mathbf{P}^{(e)}_{Q}\label{eq:q-adapt}
\end{equation}
\begin{equation}
\widetilde{\mathbf{K}}=\mathbf{K}+\sum_{e=1}^{E}\pi_e(\mathbf{c})(\mathbf{X}\mathbf{U}^{(e)}_{K})\mathbf{P}^{(e)}_{K}
\label{eq:k-adapt}
\end{equation}
Only \(\mathbf{Q}\) and \(\mathbf{K}\) were cognition-gated, as they determine inter-regional alignment, while \(\mathbf{V}\) retained task-invariant message content shared across subjects.

\textit{Phase–SC biased graph attention.} Attention logits integrate content similarity with SC and phase-based priors:
\vspace{-5pt}
\begin{equation}
\Gamma_{ij}=\frac{\langle \widetilde{\mathbf{Q}}_{i},\widetilde{\mathbf{K}}_{j}\rangle}{\sqrt{d_h}}
+\beta\,\cos(\vartheta_i^{(T)}-\vartheta_j^{(T)})
+\gamma\,\widetilde{S}_{ij},
\label{eq:plgat-gamma}
\end{equation}
where \(\beta,\gamma\in\mathbb{R}\) are learnable coefficients and \(\widetilde{\mathbf{S}}\) is the row-normalized SC derived from the anatomical mask \(\mathbf{A}_{\text{SC}}\). Anatomical validity was enforced by setting \(\Gamma_{ij}=-\infty\) when \([\mathbf{A}_{\text{SC}}]_{ij}=0\), yielding masked global attention weights
\(
\boldsymbol{\alpha}=\softmax(\Gamma\odot\mathbf{A}_{\text{SC}}),
\)
where \(\odot\) denotes elementwise masking. The graph-transformer update was computed globally as: 
\(
\mathbf{X}' = \boldsymbol{\alpha}\mathbf{V};
\)
propagating information across structurally defined edges while modulating interaction strengths by phase synchrony and cognition. Multi-head outputs were concatenated and linearly projected, followed by pre-normalization, residual aggregation, and a position-wise feedforward network (FFN) as : $\mathbf{Z}^{(g)} = \mathrm{FFN}\!\left(\mathrm{LayerNorm}\!\left(\mathbf{X}' + \mathbf{X}\right)\right)$; yielding updated node representations \(\mathbf{Z}^{(g)}\in\mathbb{R}^{N\times d_x}\). Stacking \(G\) such graph-transformer layers with shared topology \(\widetilde{\mathbf{S}}\) produced the final embedding \(\mathbf{Z}=\mathbf{Z}^{(G)}\), that encoded globally integrated yet anatomically constrained representations.

\vspace{-10pt}
\subsection{Joint Training Objective}
\vspace{-5pt}
From the final encoder output \(\mathbf{Z}\), a mean-pooling operation was applied across all nodes to obtain a graph-level embedding
\(\mathbf{u}=\frac{1}{N}\sum_{i=1}^{N}\mathbf{Z}_i\)
which served as the global representation used for downstream prediction tasks.

A sigmoid classifier predicted \(\hat{y}=\sigma(\mathbf{w}^\top\mathbf{u}+b)\), and the classification loss employed focal weighting as $\mathcal{L}_{\text{cls}}=-\eta_1(1-\hat{y})^{\gamma_f}\log(\hat{y})-\eta_0\hat{y}^{\gamma_f}\log(1-\hat{y})$, where \(\eta_0,\eta_1>0\) balance class priors and \(\gamma_f\) controls focusing strength.

In order to align learned embeddings with empirical FNC, the model reconstructed a predicted FNC as \(\widehat{\mathbf{F}}=\tfrac{1}{2}(\mathbf{Z}\mathbf{Z}^\top+\mathbf{Z}^\top\mathbf{Z})\), where pairwise inner products of node embeddings approximate inter-network co-activation. The symmetrized form ensures numerical consistency with the empirical FNC. The upper-triangular entries were vectorized as \(\mathbf{f}_{\text{pred}}\) and compared with the empirical \(\mathbf{f}_{\text{true}}\); the reconstruction loss minimizes their complement correlation as $\mathcal{L}_{\text{rec}} = 1 - \mathrm{corr}(\mathbf{f}_{\text{pred}}, \mathbf{f}_{\text{true}})$.

To overcome class imbalance problem in large scale cohort a supervised contrastive objective \cite{M_3} was applied in a normalized projection space \(\mathbf{z}=g(\mathbf{u})/\|g(\mathbf{u})\|_2\), encouraging intra-class compactness and inter-class divergence as: $\mathcal{L}_{\text{con}} = -\sum_{n}\frac{1}{|\Psi(n)|} \sum_{p\in\Psi(n)} \log \frac{\exp(\mathbf{z}_n^\top\mathbf{z}_p/\tau)} {\sum_{a\neq n}\exp(\mathbf{z}_n^\top\mathbf{z}_a/\tau)}$; where \(\Psi(n)\) denotes the set of samples sharing the same label as \(n\), and \(\tau>0\) is the temperature parameter.

The overall training objective can be formulated as:
\vspace{-5pt}
\begin{equation}
\mathcal{L}_{\text{total}}=\mathcal{L}_{\text{cls}}+\phi_{\text{rec}}\mathcal{L}_{\text{rec}}+\phi_{\text{con}}\mathcal{L}_{\text{con}},
\label{eq:joint-loss}
\vspace{-5pt}
\end{equation}
where \(\phi_{\text{rec}},\phi_{\text{con}}>0\) weight the contribution of each term. 

\vspace{-11pt}
\section{Results and Discussion}
\vspace{-8pt}
\subsection{Dataset and Pre-processing}
\vspace{-5pt}
We employed data from the Adolescent Brain Cognitive Development (ABCD) study \cite{D_0}, a large U.S. multisite longitudinal cohort spanning 21 sites. Baseline data (ages $9$–$10$) were used, including $10,199$ participants with available rs-fMRI, DTI, and cognitive assessments, where $601$ individuals were identified as having prenatal exposure to cannabis.

Resting-state fMRI data underwent standard preprocessing procedures involving slice-timing correction, motion realignment, spatial normalization, and smoothing. Intrinsic connectivity networks (ICNs; $N=53$) were derived using the \textit{NeuroMark} \cite{D_1} group ICA framework. The corresponding ICN time courses were band-pass filtered ($0.01$–$0.15$\,Hz) and z-normalized. FNC matrices $(53\times53)$ were then computed using normalized Dynamic Time Warping (nDTW) \cite{D_2} across ICN pairs. The nDTW window length was selected according to the low-frequency cutoff, resulting in 110 temporal samples for a TR of 0.8\,s.

For SC, diffusion tensors were estimated from DTI via FSL, followed by whole-brain deterministic tractography in CAMINO. Streamlines were generated in each subject’s native space, then normalized to the  \textit{NeuroMark} atlas. Fractional anisotropy maps were spatially registered to MNI space using ANTs. The resulting SC matrices $(53\times53)$ represented fiber counts between  \textit{NeuroMark}-defined networks \cite{base4}.
\vspace{-13pt}
\subsection{Experimental Settings}
\vspace{-5pt}
KOCOBrain was trained under a 5-fold stratified cross validation setup using PyTorch on NVIDIA V100 GPUs. Optimization used AdamW (learning rate $5\times10^{-4}$, weight decay $3\times10^{-4}$) for 80 epochs with batch size 16 and dropout 0.4. A grid search identified the optimal configuration with $G=3$ graph-transformer layers (6 attention heads, $d_x=96$), cognition-gate dimension $d_g=192$, and Top-$k=5$ structural masking. The Kuramoto module evolved over $T=16$ integration steps ($\Delta t=0.06$) with phase noise 0.005, while cognition gating used three low-rank experts (rank $r_e=32$). Training employed the joint objective with loss weights $\phi_{\text{rec}}=0.6$ and $\phi_{\text{con}}=0.1$.
\vspace{-11pt}
\subsection{Quantitative Evaluation} 
\vspace{-5pt}
\textit{Baselines.} We benchmarked KOCOBrain against representative SOTA families that either model graphs without explicit SC-FC coupling (GAT \cite{base1}, Graph Transformer \cite{base2}), perform generic graph learning on multimodal inputs (GCNN \cite{base3}), or fuse modalities via joint/contrastive embeddings (Joint DCCA \cite{base4}, Joint GCN \cite{base5}, BrainNN \cite{base6}). All methods used identical inputs and five-fold splits for a fair comparison. As shown in Table~\ref{Table_1}, GAT and Graph Transformer provided global attention but treated SC and FNC as static features; lacking phase-informed routing or anatomical gating, they performed ${\sim}11$–$13$ pp lower in accuracy (acc.), ${\sim}11$–$12$ pp in specificity (spec.), and notably less (${\sim}27$–$29$ pp) in sensitivity (sens.) compared to KOCOBrain. GCNN and joint fusion baselines integrated modalities but lacked dynamics and cognition-aware modulation, resulting in comparatively lower performance than ours. The strongest competitor, BrainNN, still underperformed KOCOBrain by $5.2$ pp acc., $11.6$ pp sens., and $4.8$ pp spec., indicating that Kuramoto-guided structure-function coupling plus cognition-gated attention produced more discriminative and biologically consistent representations for PDE prediction. 

\textit{Ablations.} As listed in Table~\ref{Table_2}, ablations were organized around 3 key research questions (RQs) examining core model components. \textit{RQ1–Kuramoto-based coupling:} Removing the phase synchronization dynamics ($\boldsymbol{\vartheta}^{(T)}$) led to a substantial drop in acc to 76.13\%, highlighting the necessity of dynamic multimodal coupling for coherent representation learning. Excluding phase or structural bias terms ($\beta \cos(\Delta\vartheta)$, $\gamma\,\widetilde{S}$) similarly degraded performance, indicating that both synchrony and anatomical weighting jointly stabilized attention under physiological constraints. \textit{RQ2–Cognition-gated attention:} Excluding cognitive context $(c)$ or disabling the MoE adapters consistently reduced sensitivity by ${\sim}3.3$ pp, suggesting that subject-specific gating facilitated adaptive information routing across individuals with variable cognitive profiles. \textit{RQ3–Multi-objective optimization:} Disabling auxiliary objectives weakened both robustness and discriminability; removing reconstruction term $(\phi_{rec})$ reduced alignment with empirical FNC, while omitting the contrastive term $(\phi_{con})$ led to poorer class separation under imbalance. 

\begin{table}[t]
\caption{Comparative evaluation against leading state-of-the-art (SOTA) baselines (Mean $\pm$ Standard Deviation).}
\vspace{-5pt}
\begin{center}
\resizebox{\columnwidth}{!}{\begin{tabular}{|c|c|c|c|} 
\hline
\textbf{Method} & \textbf{Accuracy} & \textbf{Sensitivity} & \textbf{Specificity} \\
\hline
GAT \cite{base1} & 0.7045 $\pm$ 0.0049 & 0.5341 $\pm$ 0.0183 & 0.7152 $\pm$ 0.0043 \\ \hline
Graph Transformer \cite{base2} & 0.7102 $\pm$ 0.0055 & 0.5554 $\pm$ 0.0330 & 0.7199 $\pm$ 0.0074 \\ \hline
GCNN  \cite{base3} & 0.6942 $\pm$ 0.0153 & 0.5325 $\pm$ 0.0690 & 0.7044 $\pm$ 0.0204 \\ \hline
Joint DCCA \cite{base4} & 0.7630 $\pm$ 0.0049 & 0.7018 $\pm$ 0.0088 & 0.7668 $\pm$ 0.0057 \\ \hline
Joint GCN \cite{base5} & 0.7664 $\pm$ 0.0064 & 0.6749 $\pm$ 0.0301 & 0.7721 $\pm$ 0.0053 \\ \hline
BrainNN \cite{base6} & 0.7849 $\pm$ 0.0101 & 0.7155 $\pm$ 0.0235 & 0.7892 $\pm$ 0.0118 \\ \hline
\textbf{KOCOBrain} & \textbf{0.8365 $\pm$ 0.0018} & \textbf{0.8316 $\pm$ 0.0051} & \textbf{0.8368 $\pm$ 0.0022} \\
\hline

\end{tabular}}
\vspace{-25pt}
\label{Table_1}
\end{center}
\end{table}

\begin{table}[t]
\caption{Comparative performance of KOCOBrain variants under ablation settings (Mean $\pm$ Standard Deviation).}
\vspace{-15pt}
\begin{center}
\resizebox{\columnwidth}{!}{\begin{tabular}{|c|c|c|c|} 
\hline
\textbf{KOCOBrain Variants} & \textbf{Accuracy} & \textbf{Sensitivity} & \textbf{Specificity} \\
\hline
w/o Kuramoto dynamics: \(\boldsymbol{\vartheta}^{(T)}\) & 0.7613 $\pm$ 0.0040 & 0.7331 $\pm$ 0.0234 & 0.7631 $\pm$ 0.0053 \\ \hline
w/o Phase bias: $\beta \cos(\Delta\vartheta)$ & 0.7778 $\pm$ 0.0025 & 0.7577 $\pm$ 0.0036 & 0.7791 $\pm$ 0.0029 \\ \hline
w/o SC bias: $\gamma\,\widetilde{S}$ & 0.8118 $\pm$ 0.0023 & 0.7641 $\pm$ 0.0048 & 0.8148 $\pm$ 0.0026 \\ \hline
w/o cognition: $c$ & 0.8066 $\pm$ 0.0095 & 0.7990 $\pm$ 0.0130 & 0.8070 $\pm$ 0.0093 \\ \hline
w/o MoE adapters & 0.8272 $\pm$ 0.0010 & 0.8003 $\pm$ 0.0024 & 0.8289 $\pm$ 0.0012 \\ \hline
w/o FC recon: $\phi_{\text{rec}}$ & 0.8115 $\pm$ 0.0015 & 0.7980 $\pm$ 0.0050 & 0.8123 $\pm$ 0.0019 \\ \hline
w/o SupCon: $\phi_{\text{con}}$ & 0.8209 $\pm$ 0.0012 & 0.7607 $\pm$ 0.0029 & 0.8246 $\pm$ 0.0013 \\ \hline
\textbf{Proposed} & \textbf{0.8365 $\pm$ 0.0018} & \textbf{0.8316 $\pm$ 0.0051} & \textbf{0.8368 $\pm$ 0.0022} \\
\hline

\end{tabular}}
\vspace{-25pt}
\label{Table_2}
\end{center}
\end{table}
\vspace{-11pt}
\subsection{Qualitative Evaluation} 
\begin{figure}[htb]
  \centering
  \centerline{\includegraphics[width= 6 cm,height=3 cm]{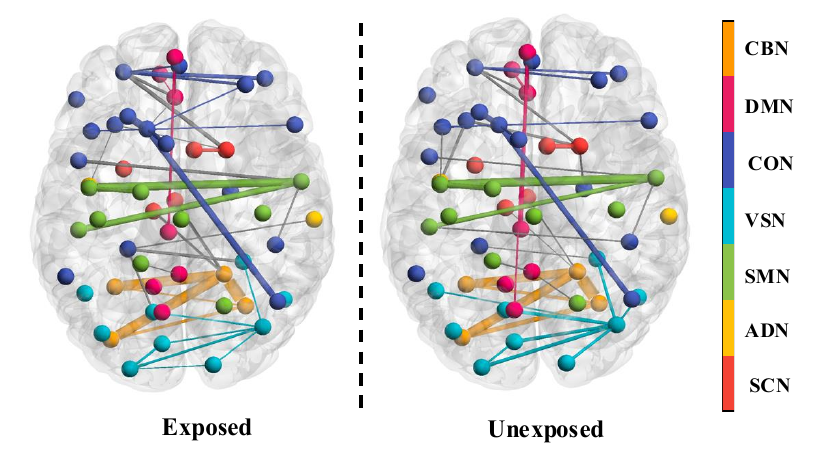}}
  \vspace{-10pt}
\caption{Axial depiction of group-mean attention maps highlighting the top 3\% connections for exposed and unexposed groups across \textit{NeuroMark} networks: subcortical (SCN), auditory (ADN), sensorimotor (SMN), visual (VSN), cognitive control (CON), default mode (DMN), and cerebellar (CBN) network. Intra-network links are color-coded, while inter-network edges appear in gray, with line thickness scaled to attention strength to indicate relative connection importance.
}
\vspace{-15pt}
\label{fig:fig2}
\end{figure}

\vspace{-5pt}
To identify the network interactions most discriminative for PDE prediction, we examined the top 3\% of attention weights $\boldsymbol{\alpha}$ averaged across subjects within each group and illustrated in Fig. \ref{fig:fig2}. These values capture how structurally constrained networks communicate under the joint influence of phase synchrony and cognition. In unexposed adolescents, strong attentional coupling was observed among the SMN, VSN, and CON, extending toward the DMN--a distributed integration pattern typical of normative large-scale coordination \cite{I_0}. In contrast, the exposed group exhibited more focal organization, with dominant connections linking the DMN, CON, and CBN. This concentrated pattern aligns with recent findings \cite{Qual_1,Qual_3,My_1} showing that PDE disrupts global integration and enhances coupling within executive, salience, and cerebellar circuits associated with cognitive control. Together, these patterns suggest that PDE shifts brain organization toward selective engagement of higher-order control circuits, reflecting adaptive reconfiguration during development.

\vspace{-10pt}
\section{Conclusions}
\vspace{-10pt}
We introduced KOCOBrain, a unified framework that bridges structural and functional connectomes via  Kuramoto-guided dynamics and cognition-gated graph attention. By coupling oscillatory synchronization with subject-specific modulation, it achieves robust and interpretable prediction of PDE status. The integration of dynamic coupling, cognition-aware attention, and multi-objective optimization enhanced model sensitivity and stability when applied to the imbalanced PDE prediction task within the ABCD cohort. Future work will extend this framework to longitudinal settings to investigate developmental trajectories and evolving brain-behavior associations.

\section{Compliance with Ethical Standards}
\vspace{-8pt}
This study utilized existing human subject data in accordance with all applicable ethical guidelines.
\vspace{-10pt}

\bibliographystyle{IEEEbib}
\bibliography{refs}

\end{document}